\documentclass[twocolumn,groupedaddress,superscriptaddress,lengthcheck]{aastex631}
\usepackage{amsmath}
\usepackage{float}
\usepackage{hyperref}
\usepackage{xspace}
\usepackage{natbib}

\newcommand{\m}{M\textsubscript{\(\odot\)}\xspace}

\begin{document}

\title{Mining the Alerts: A Preliminary Catalog of Compact Binaries from the Fourth Observing Run}

\author[0000-0001-7248-951X]{Aleyna Akyüz}
\affiliation{Department of Physics, Syracuse University, Syracuse, NY 13244, USA}

\author[0009-0000-9756-9794]{Alex Correia}
\affiliation{Department of Physics, Syracuse University, Syracuse, NY 13244, USA}

\author[0009-0001-5699-2294]{Jada Garofalo}
\affiliation{Department of Physics, Syracuse University, Syracuse, NY 13244, USA}

\author[0000-0001-8051-7883]{Keisi Kacanja}
\affiliation{Department of Physics, Syracuse University, Syracuse, NY 13244, USA}

\author[0009-0008-5970-9694]{Vikas Jadhav Y}
\affiliation{Department of Physics, Syracuse University, Syracuse, NY 13244, USA}

\author[0009-0004-8445-6212]{Labani Roy}
\affiliation{Department of Physics, Syracuse University, Syracuse, NY 13244, USA}

\author[0000-0001-8051-7883]{Kanchan Soni}
\affiliation{Department of Physics, Syracuse University, Syracuse, NY 13244, USA}

\author[0000-0001-9101-048X]{Hung Tan}
\affiliation{Department of Physics, Syracuse University, Syracuse, NY 13244, USA}

\author[0000-0002-0355-5998]{Collin D. Capano}
\affiliation{Department of Physics, Syracuse University, Syracuse, NY 13244, USA}
\affiliation{Physics Department, University of Massachusetts Dartmouth, North Dartmouth, MA 02747, USA}

\author[0000-0002-1850-4587]{Alexander H. Nitz}
\affiliation{Department of Physics, Syracuse University, Syracuse, NY 13244, USA}

\correspondingauthor{Alexander Harvey Nitz}
\email{ahnitz@syr.edu}

\begin{abstract}
We present a preliminary catalog of compact binary merger candidates from the ongoing fourth observing run (O4) of Advanced LIGO, Virgo, and KAGRA, based on an analysis of public alerts distributed through GraceDB as of May 2025. We developed and applied methods to estimate the source-frame chirp mass for each candidate by utilizing information from public data products, including source classification probabilities, sky localizations, and observatory status. Combining our O4 analysis with previous catalogs, we provide updated estimates for the local merger rate density. For sources with chirp mass characteristic of binary neutron stars ($[1, 1.5]\,\m$), we find a rate of $56^{+99}_{-40}$~$\textrm{Gpc}^{-3}\,\textrm{yr}^{-1}$. For systems in the expected neutron star--black hole chirp mass range ($[1.5, 3.5]\,\m$), the rate is $36^{+32}_{-20}$~$\textrm{Gpc}^{-3}\,\textrm{yr}^{-1}$, and for heavier binary black holes ($[3.5, 100]\,\m$), we estimate a rate of $19^{+4}_{-2}$~$\textrm{Gpc}^{-3}\,\textrm{yr}^{-1}$. This work provides an early glimpse into the compact binary population being observed in O4; we identify a number of high-value candidates up to signal-to-noise $\sim 80$, which we expect to enable precision measurements in the future.
\end{abstract}

\section{Introduction} \label{sec:intro}

The fourth observing run (O4) of the Advanced LIGO~\citep{LIGOScientific:2014pky}, Virgo~\citep{VIRGO:2014yos}, and KAGRA~\citep{KAGRA:2018plz} gravitational-wave observatories is poised to be the most prolific to date. Building on the foundation of previous runs, which have produced a catalog of nearly one hundred compact binary mergers~\citep{KAGRA:2021vkt,  Nitz:2021zwj,PhysRevD.106.043009, Mehta:2023zlk, Wadekar:2023gea}, O4 is expected to more than triple the number of observed gravitational-wave observations. This rapid accumulation of data provides an unprecedented opportunity to refine our understanding of the compact object mass spectrum, the cosmic merger rate history, and the formation channels of binary systems~\citep{LIGOScientific:2020ufj,Mandel:2021smh,Gerosa:2021mno,Edelman:2021fik,Zevin:2020gbd,KAGRA:2021duu}.

The growing gravitational-wave catalog has already revealed key features of the underlying compact object population~\citep{KAGRA:2021duu}, including possible signs of precessing binaries~\citep{Hannam:2021pit,2025arXiv250309773H} and residual eccentric orbits~\citep{Gayathri:2020coq,Romero-Shaw:2022xko,Morras:2025xfu,Planas:2025plq,Romero-Shaw:2025vbc}. We have observed a continuum of binary black hole (BBH) mergers along with a handful of neutron star binary mergers~\citep{TheLIGOScientific:2017qsa, Abbott:2020uma} and neutron star black hole (NSBH) mergers~\citep{LIGOScientific:2021qlt}. These observations have begun to probe the masses between neutron stars and black holes and have begun to populate the upper mass gap~\citep{Mangiagli:2019sxg,LIGOScientific:2020iuh,LIGOScientific:2020ufj,Nitz:2020mga,Ezquiaga:2020tns,Tanikawa:2020cca,Mehta:2021fgz}, previously thought to be devoid of black holes due to pair-instability supernovae~\citep{Barkat:1967zz,Bond:1984sn,Woosley:2007qp,Woosley:2021xba,Umeda:2001kd}.

% summary of the method, what's new, set expecations appropriately
\begin{figure*}
    \centering
    \includegraphics[width=1\textwidth]{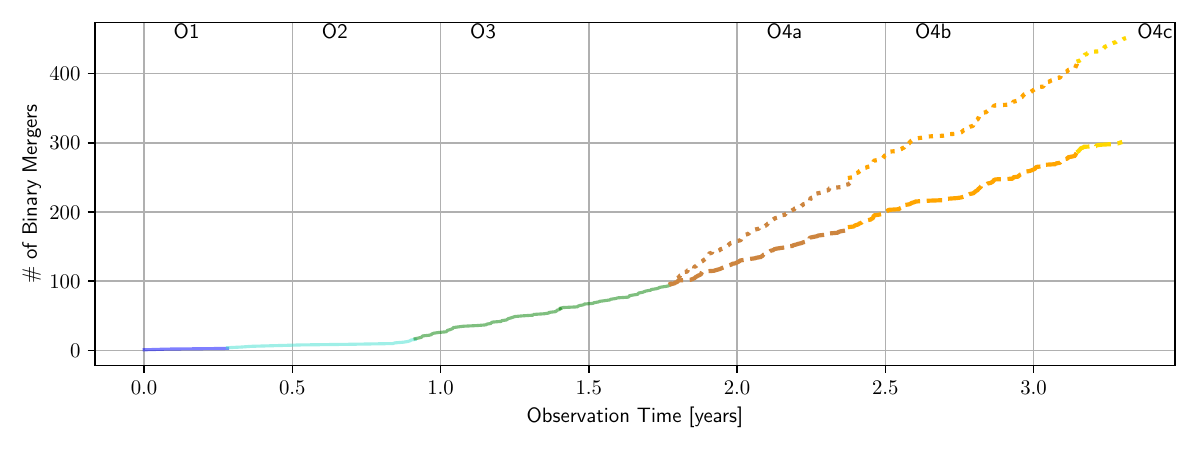}
    \caption{The number of compact binary merger observations as a function of observing time. For the first three observing runs (O1-O3) these are sourced from the 4-OGC catalog and for O4 they are sourced from the  Gravitational-Wave Candidate Event Database (GraceDB) using a false alarm rate threshold of $10^{-7}$Hz (dashed) or $10^{-6}$Hz (dotted). The O1-O4 observing runs are shown (various colors). The O4 run is split into three parts O4a (brown), O4b (orange), and the ongoing run O4c (yellow). The most substantial improvements in relativity sensitivity occurred between observing runs O2 and O3, hence the significant change in slope; while there were improvements in detector sensitive range between O3 and O4, they were modest in comparison ($\sim 20-30\%$). However, in part due to the substantial increase in observing time, O4 will contain more observations than all prior observing runs combined; O4a alone should double the number of known sources. 
    }
    \label{fig:observing}
\end{figure*}

The LIGO, Virgo, and KAGRA (LVK) collaborations release key information about candidate events in low latency with the goal that this information is useful for follow-up efforts~\citep{gracdb_dcc}. The best case of this is evidenced by GW170817~\citep{LIGOScientific:2017vwq}, a binary neutron star merger, which was observed in coincidence with a gamma-ray burst~\citep{LIGOScientific:2017zic}, and whose prompt localization enabled the discovery of the associated kilonova~\citep{LIGOScientific:2017ync}. While binary black hole mergers are not typically expected to produce electromagnetic counterparts, they might if they merge in a matter-rich environment~\citep{Bartos:2016dgn}. One such scenario is a merger within the disk of an active galactic nucleus (AGN); this has been proposed for GW190521, which had a potential coincident flare observed by the Zwicky Transient Facility~\citep{Graham:2020gwr}.

In this paper, we present an analysis of public alerts from the ongoing fourth observing run to provide a preliminary catalog and population preview. We utilize publicly available low-latency data products, including source classification probabilities and sky localizations, to derive estimates of the source-frame chirp mass for each candidate. By combining our O4 estimates with the established catalog from previous runs, we analyze the growing population of compact binaries and provide updated constraints on their merger rates.

% comment on some of the key results / summary

\section{Determining Source Properties}

The LIGO, Virgo, and KAGRA (LVK) scientific collaborations release information about public candidates through the GraceDB server~\citep{gracdb_dcc,KAGRA:2023pio,Chaudhary:2023vec}. These public alerts include estimates of the source location~\citep{Singer:2015ema} and source classification~\citep{Kapadia:2019uut,Chatterjee:2019avs,Andres:2021vew,Villa-Ortega:2022qdo}; they are made available to aid telescopes in the search for potential counterparts. By combining this low-latency information with the known sensitivity of the observatories, it is possible to determine key properties of each observation, including a candidate's signal-to-noise ratio (SNR) and its chirp mass. In this work, we detail how we combine three complementary approaches to provide these estimates for each source observed to date in O4: (1) using the relative source classification probabilities, (2) fitting the expected signal SNR as a function of mass to the observed distance and detector sensitivity, and (3) comparing the sky localization shape.

% This kind of thing needs to be somewhere? 
We note that while one can obtain accurate estimates of a candidate's chirp mass with these methods, it is less straightforward to obtain information about its mass ratio or spin properties. These characteristics hold a wealth of information to determine the formation history of the overall population~\citep{KAGRA:2021duu} and will be a focus of intense investigation as more detailed data becomes available.

\subsection{Source Classification}

\begin{table}
\centering
\includegraphics[width=\columnwidth]{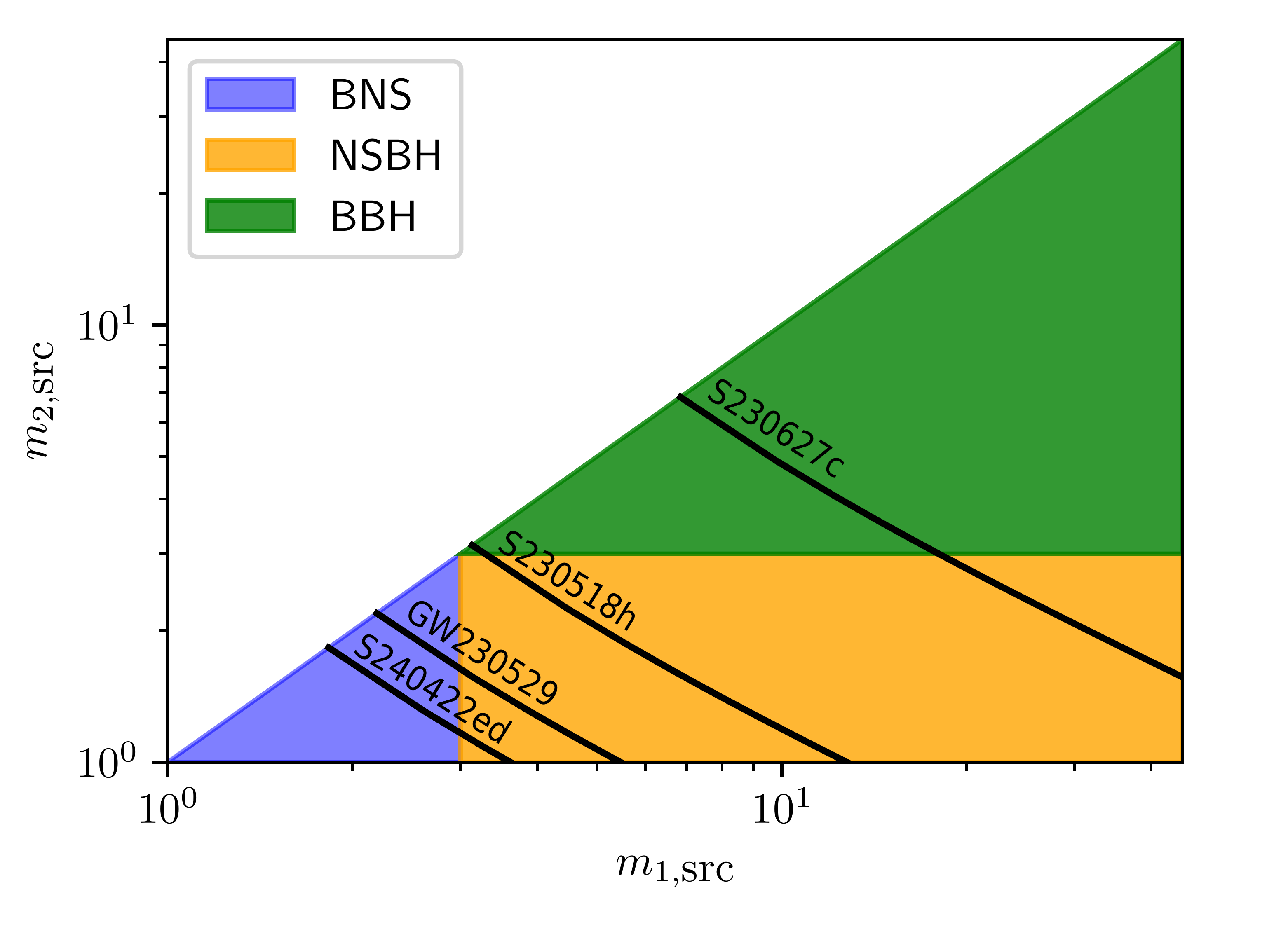}
\begin{tabular}{|l|l|l|l|l|}
\hline
 Candidate   & BNS & NSBH & BBH  &  $\mathcal{M}_{\textrm{src}}$ [$M_{\odot}$]\\\hline
  S230518h  & 0 & 0.959  & 0.041  &  $2.77^{+.04}_{-.04}$  \\
 GW230529 & 0.329 & 0.671 & 0  &  $1.93^{+.04}_{-.05}$  \\
  S230627c & 0 & 0.493 & 0.507  &  $6.04^{+.10}_{-.11}$  \\
  S240422ed & 0.700 & 0.300  & 0  &  $1.61^{+.02}_{-.02}$  \\
  \hline
\end{tabular}
\caption{
%Estimation of the chirp mass $\mathcal{M}_{\textrm{src}}$ of a compact-binary candidate based on their astrophysical origin source classification probabilities from the \textit{PyCBC Live} public results.
A schematic picture of how the source-frame chirp mass estimate results in a source classification (top figure) for a few sample candidates along with the public source probabilities from the \textit{PyCBC Live} public results and estimated source-frame chirp mass (bottom table). Uncertainties are given as at the $90\%$ credible level. Notably, the estimate for GW230529 is consistent with the published value of $1.94^{+.04}_{-.04} M_{\odot}$~\citep{LIGOScientific:2024elc}.}
\label{fig:mchirptable}
\end{table}

The chirp mass of a compact-binary candidate can be estimated by careful interpretation of the low-latency alert products. Several low-latency compact binary searches are based on matched filtering against a bank of waveform templates~\citep{DalCanton:2020vpm,Sachdev:2019vvd,Andres:2021vew,Hooper:2011rb}. For each candidate, identifying the template directly provides a point estimate of the source's detector-frame (redshifted) chirp mass that is already within 1\% of the true value for BNS and NSBH sources~\citep{Villa-Ortega:2022qdo}.

The method of~\citet{Villa-Ortega:2022qdo}, used by the PyCBC Live~\citep{DalCanton:2020vpm} and SPIIR~\citep{Hooper:2011rb} pipelines, creates a deterministic mapping between the source-frame chirp mass of a candidate and its source classification probabilities. These probabilities assume: (1) a mass distribution that is uniform in the source-frame component masses, and (2) a well-measured detector-frame chirp mass. Since we have the resulting source classification estimates and a source's distance in low latency, we can reverse this process to obtain an estimate of the source-frame chirp mass. The chirp mass is estimated using a $\chi^2$ minimization of the predicted-to-observed source classification probabilities, achieving a relative tolerance of $<10^{-8}$. The uncertainty in each estimate is derived from the uncertainty on the distance (via the redshift scaling factor) used to convert from detector- to source-frame chirp mass; sub-dominant systematic uncertainties on the detector-frame chirp mass can be ignored. How this process works is illustrated in Table.~\ref{fig:mchirptable}. This method is effective when a source classification of this type is available and when the probabilities are not entirely within a single class. This restricts the approach to sources with a source-frame chirp mass $\mathcal{M} < \sim 9~ M_\odot$.

\subsection{Skymap-based Estimation}

\begin{figure*}
    \centering
    \includegraphics[width=0.48\textwidth]{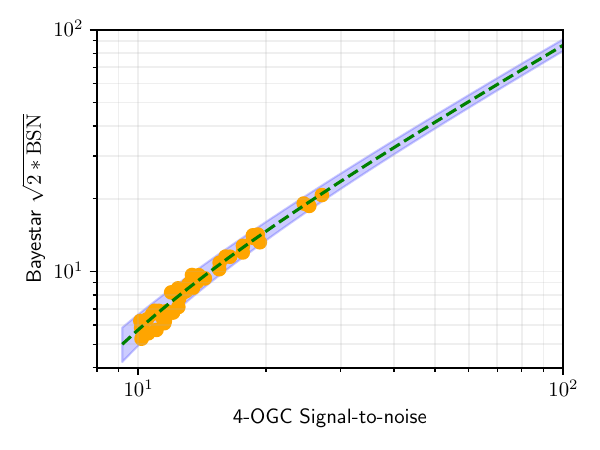}
    \includegraphics[width=0.48\textwidth]{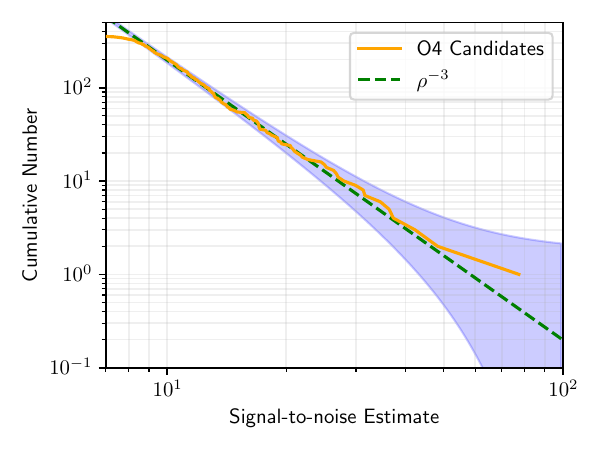}
    \caption{(Left) The signal-to-noise (SNR) of known gravitational-wave signals from the 4-OGC catalog compared to the Bayestar-derived BSN (orange). There is a nearly linear relationship between the SNR and $\sqrt{2 * \textrm{BSN}}$; the correlation coefficient $r=0.993$. The best fit line (green dashed) along with the 2-sigma standard deviation in the $\chi^2$ fit to the data (blue) are shown. (Right) The distribution of signal-to-noise ratio for observed candidates in O4 with false alarm rate $< 10^{-6}$Hz. For comparison, the expected distribution for a population of sources that are distributed uniformly in luminosity volume is shown (green) along with the expected 1-sigma uncertainties in the expected number assuming a Poisson distribution (blue). We see that the cumulative distribution of candidate SNR closely matches the source distribution expectation; a deviation at low SNR ($<10$) is expected due to the extinction of sources both from the imposed FAR threshold and from searches failing to detect parts of the population. 
    }
    \label{fig:snrfit}
\end{figure*}

We employ two methods to help determine the chirp mass of a source using information from the public skymaps. The first works by comparing the expected SNR for a given chirp mass with the SNR derived from the skymap. The second directly compares the shape of the localization posterior to those from sources with different properties. The former method will be used primarily in this work as it yields lower uncertainties, while the latter will be employed to help resolve ambiguities in the estimates from the other methods.

For each source, a BAYESTAR-produced skymap is available~\citep{Singer:2015ema}. The skymap includes the posterior distribution on sky location and distance, along with an estimate of the Bayesian signal-to-noise (BSN)~\citep{Singer:2015ema}. Since the BSN is related to the coherent network SNR, $\rho$, we can directly obtain an estimate of the signal's SNR. For approximately 40 observations in O1-O3, we can fit the obtained BSN to the measured SNR from detailed analyses. As shown in Fig.~\ref{fig:snrfit}, we find that $\rho^2$ and $\ln(BSN)$ have a highly linear relationship (r=0.993). This fit can then be used to estimate the SNR of all remaining sources. 

To make this method work, we determine the sensitivity of the instruments and fix all properties of the source that could affect its SNR, leaving only the chirp mass as a free parameter to be fitted. This procedure is performed for the detector-frame masses, which are later converted to the source frame using distance estimates. We determine the sensitivity of each instrument at the time of observation from the public observatory status pages~\citep{KAGRA:2023pio}\footnote{This website is accessible at \url{https://gwosc.org/detector_status/}}, which provide both the average shape of the noise curve and the specific sensitive range of the instrument at that time.

We assume that the mass ratio and component spins of each source are close to unity and zero, respectively. These properties are true for the vast majority of sources~\citep{KAGRA:2021duu}, but this does present a potential source of systematic bias. Notably, deviations from these assumptions only incur a systematic bias when they affect the SNR of a signal.

The sky location and distance of a source are assumed to be at the maximum a posteriori (MAP) position. This leaves only two extrinsic parameters that have the potential to significantly affect the expected SNR: the binary inclination and the polarization angle. In most cases, both of these parameters are poorly measured~\citep{Usman:2018imj,KAGRA:2021duu}; instead, only an overall amplitude is well constrained. The average value of the inclination for the peak in the distance distribution can be estimated from a prior constraint that requires a consistent amplitude; this gives an expected peak that corresponds to an inclination of $\sim0.6$ radians. We do not assume this is the actual inclination of the binary but rather make the significantly weaker assumption that the MAP distance corresponds to this effective inclination. 

With all other parameters now fixed, we can directly calculate the SNR that would be expected for each candidate as a function of its chirp mass. We fit the estimated SNR to its respective chirp-mass curve. There are two chirp mass solutions for every SNR, with the latter typically corresponding to a high enough mass binary such that its characteristic frequencies begin to move out of the sensitive band of the observatory. We break the degeneracy between these two solutions by comparing the respective predicted sky localizations to the observed Bayestar skymap; in all cases, the lower mass solution is visually consistent with the observed localization region. 

% discuss what information is already known and the general premise
\begin{figure}
    \centering
    \includegraphics[width=0.5\textwidth]{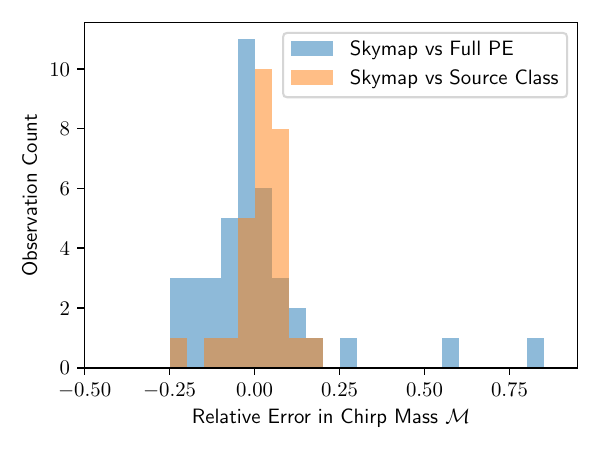}
    \caption{
    The relative difference in chirp mass between the published values from full parameter estimation and our skymap-informed method (blue) and between our source and skymap derived methods (orange). Several outliers exist, which are associated with sources that prefer either a high spin component (positive outliers) or a highly unequal mass ratio (negative outliers). These are both cases where our model would be expected to break down. Notably, this is also more prone to occur for the highest mass binaries.
    }
    \label{fig:validate}
\end{figure}

% include validation plot
\subsection{Validation and Systematics}

To validate our chirp mass estimation procedures and quantify their systematic uncertainties, we perform two primary comparisons. First, we apply our skymap-based estimation method to well-characterized events from the O1-O3 observing runs, as cataloged in 4-OGC~\citep{Nitz:2021zwj}, and compare our derived chirp masses against the published values from full parameter estimation. Second, for the new O4 candidates, we perform an internal consistency check by comparing the estimates obtained from our skymap-based method against those from the source-classification method. This allows us to both anchor our results to known sources and ensure that our methods produce consistent results for the new population of events.

The results of these comparisons are shown in Fig.~\ref{fig:validate}, where our estimates are broadly consistent across both validation tests. The agreement between our two internal O4 methods is comparable to the agreement found when validating against the O1-O3 catalog. This result is expected, as the source-classification method is most accurate for low-mass systems, and any discrepancy is thus dominated by the systematics of the skymap method. We identify a handful of outliers, which we attribute to a breakdown in our model's simplifying assumptions. We find that the most extreme cases of bias occur for sources where detailed analyses have shown a preference for unequal mass ratios or significant component spins, such as GW190814 and GW190521~\citep{LIGOScientific:2020zkf, LIGOScientific:2020ufj}, as these properties affect signal morphology and amplitude in ways not captured by our fiducial model. Further confidence in our methods for O4 events comes from the recent analysis of the NSBH merger GW230529; our estimate of its chirp mass is in excellent agreement with the published value~\citep{LIGOScientific:2024elc}.

Even including these outliers, we find that our methods constrain the chirp mass to within a 20\% uncertainty at a 90\% confidence level. We expect that the primary source of this uncertainty stems from our simplifying assumption for the binary inclination, which directly impacts the inferred signal amplitude. However, for specific events, particularly the outliers, the breakdown of our assumptions—namely that the binaries are equal-mass and non-spinning—provides a significant contribution to the error. Finally, we cannot preclude that in some cases, the initial low-latency search pipelines may have identified a candidate with a template that is not fully representative of the true signal, introducing an additional source of error into our estimates.

There are promising avenues to mitigate these systematics. One dominant source of error, our fixed assumption for the binary inclination, could be improved by leveraging information from the distance posterior itself. The width of the distance distribution is correlated with the viewing angle, and this can be used to select a more probable effective inclination corresponding to the MAP distance, rather than adopting a single average value for all events.

\section{Population of Compact Binaries}

\begin{figure*}
    \centering
    \includegraphics[width=0.497\textwidth]{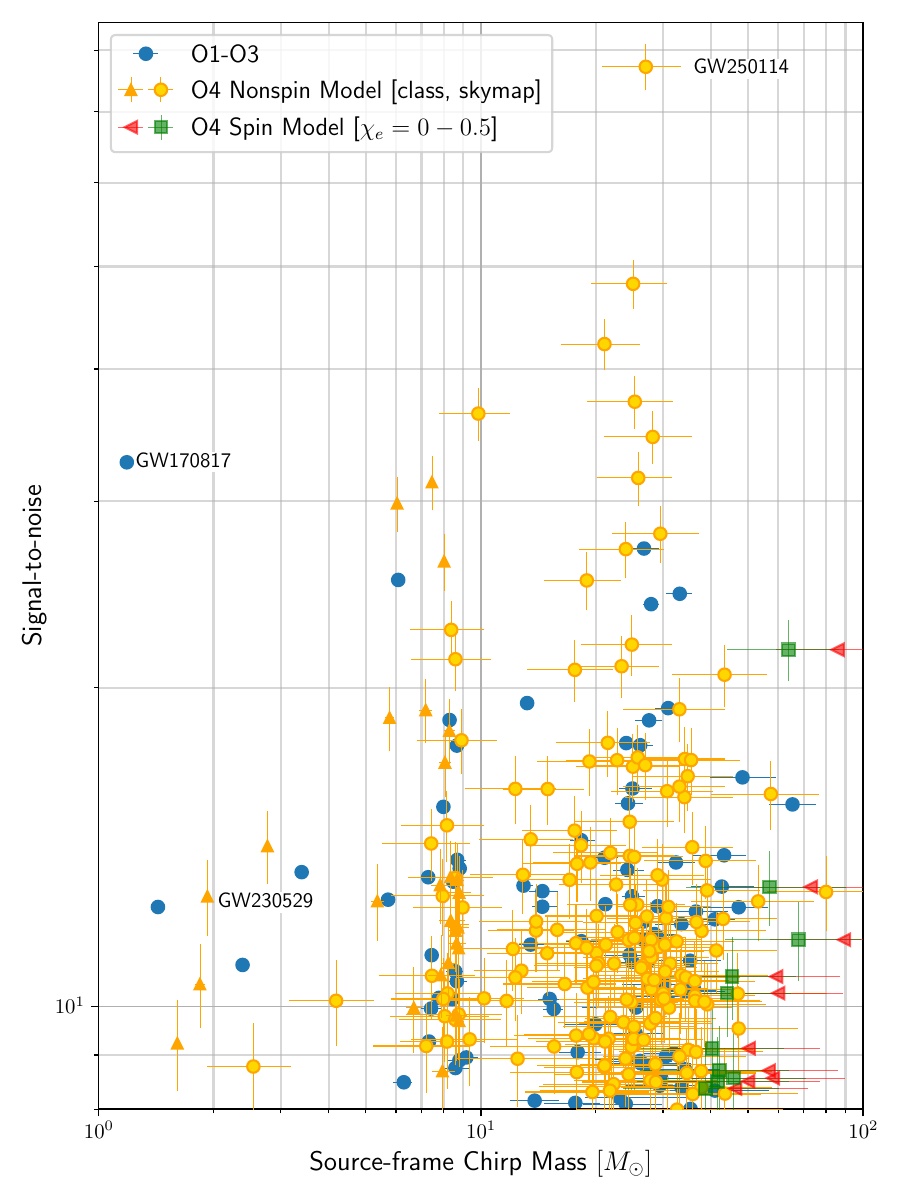}
    \includegraphics[width=0.497\textwidth]{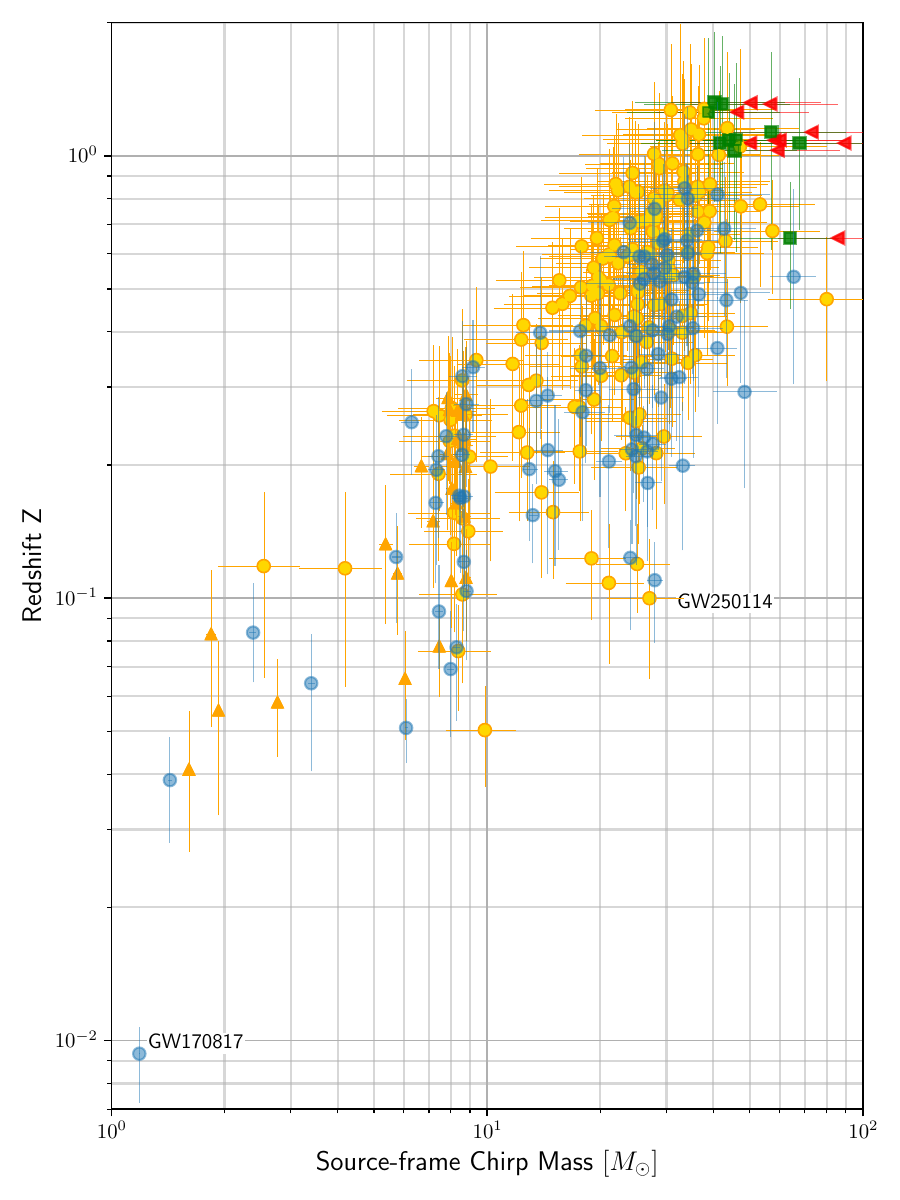}
    \caption{The population of observed compact binary mergers from O1-O4 through May 2025 as a function of source-frame chirp mass and SNR (left) or redshift z (right). Observations from O1-O3 use the values from the 4-OGC catalog (blue). Sources observed in O4 (yellow) are either derived from the source-classification (triangles) or skymap (circles). In most cases, our model assumption is that the source is composed of comparable mass objects that are nonspinning. In some cases, this nonspinning model cannot fit the observed localization information; while we cannot rule out alternate violations of our basic model, for these cases we also consider a model that allows them to have an aligned spin component up to an effective spin of 0.5 (shown in red to green). Broad agreement between the O4 observations and past O1-O3 observation is evident. There is also continued support in the O4 catalog for a sub-population of binary black hole mergers centered at a chirp mass of $\sim 13 M_\odot$.
    }
    \label{fig:pop}
\end{figure*}

The population of compact binary mergers observed in the fourth observing run is broadly consistent with that established from prior runs~\citep{KAGRA:2021duu,Nitz:2021zwj}, reinforcing key features of the underlying astrophysical distributions while substantially increasing their statistical definition. With over 200 new binary black hole mergers identified with false-alarm-rate (FAR) $< 10^{-7}$ Hz, the O4 dataset provides a clearer view of the black hole mass spectrum. Figure~\ref{fig:pop} shows the distribution of sources. The chirp mass distribution continues to show two dominant features: a broad peak centered at a chirp mass of $\sim 8\,\m$ and a second prominent peak near $30\,\m$. Our analysis also finds continued support for an intermediate feature around $13\,\m$, a characteristic that has been suggested by previous non-parametric reconstructions of the GWTC-3 catalog~\citep{Tiwari:2020otp,Edelman:2022ydv}.

The complex structure of the black hole mass spectrum points towards the contribution of multiple formation channels~\citep{Mandel:2021smh}. The lower-mass peak is generally consistent with predictions from models of isolated binary evolution in galactic fields, where stellar winds play a crucial role in shaping the final black hole masses~\citep{Belczynski:2016obo,Garcia:2021niy}. The features at higher masses, however, are more difficult to explain with this channel alone. The prominent peak around $30-35\,\m$ and the emerging population of binaries in the upper mass gap are compelling signatures of dynamical formation pathways, particularly hierarchical mergers in dense stellar environments like globular clusters or disks of AGN~\citep{Rodriguez:2019huv, McKernan:2019hqs, Fishbach:2021yvy}. Our analysis identifies several candidates with masses that fall within the upper mass gap, a region from roughly $60\,\m$ to $120\,\m$ where black holes are not expected to form from the collapse of single stars due to the effects of pair-instability supernovae~\citep{Heger:2002by, Woosley:2016hmi}. Hierarchical mergers provide a natural explanation, as the collision of two black holes with masses near the top of the first-generation peak ($\sim 30-40\,\m$) would produce a remnant that falls squarely within this gap. 

A number of notable observations stand out in the O4 catalog. Foremost among these is GW250114, a candidate with an estimated chirp mass of $\sim 27^{+6}_{-6}\,\m$ and an exceptionally high network signal-to-noise ratio of $\sim$80. An event this loud provides an unprecedented opportunity for precision measurements of its source parameters, including spin-precession and potential deviations from general relativity. A more detailed follow-up of this event is explored in a companion paper~\citep{Placeholder:GW250114}. Beyond this singular event, the O4 catalog contains a further seven sources with an SNR greater than 30, bringing the total number of such high-SNR events to eight. This growing collection of loud signals will enable detailed exploration of individual source properties and their place within the broader population.

Finally, we must add a note of caution regarding the mass estimates for some of these interesting events. As discussed in our validation section, our estimation method may have higher systematic uncertainties for sources that deviate significantly from our fiducial model. Several of the potential mass-gap candidates cannot be fit to our nonspinning, fixed-inclination model. While other systematics may be at play, if we relax these assumptions and allow for significant component spin, the inferred chirp mass for some of these events shifts into agreement with the broader binary black hole distribution. Nevertheless, even with this consideration, several candidates appear to remain robustly within the upper mass gap, strengthening the case for their astrophysical origin through hierarchical channels.

%\begin{figure*}
%    \centering
%    \includegraphics[width=1\textwidth]{pop_dist.pdf}
%    \caption{
%    }
%    \label{fig:observing}
%\end{figure*}

\subsection{Merger Rate}

We estimate the merger rate of compact-binary mergers by combining the past results from the 4-OGC catalog~\citep{Nitz:2021zwj} with our estimates of the O4 observations. We take a straightforward non-parametric approach to modeling the sensitive volume of a gravitational-wave search and the observed population's mass distribution. This methodology differs from other common approaches to population analysis; studies have largely employed hierarchical Bayesian frameworks with parametric models for the underlying distributions~\citep{KAGRA:2021duu}. Other independent analyses have used flexible, non-parametric models, such as splines or Gaussian processes, to reconstruct population features with fewer assumptions about their functional form~\citep{Tiwari:2020otp,Edelman:2022ydv}.

Given that there are multiple operating low-latency analyses with overlapping search spaces and unspecified trial factors, we use a SNR threshold to create our event sample rather than one based solely on the FAR. This provides a more uniform selection criterion across the different search pipelines. We conservatively choose a network SNR of 10, which gives us $\sim 200$ sources through May 2025. We consider all sources above this threshold, irrespective of the operating instruments at the time of the event.

We estimate the selection effects of this SNR threshold and the corresponding surveyed time-volume for O1-O4 by a simple Monte Carlo simulation. This calculation determines the expected SNR for a population of sources that is assumed to be distributed isotropically on the sky and in their orientation. We account for both the sensitivity of each instrument as a function of time and their observational status using information collected from prior catalogs for O1-O3 and from the public observatory summary pages for O4. As the O4 strain data is not yet public, we obtain the instrumental sensitivity curves for each day of operation, along with the higher-cadence sensitive range estimates, from these public data streams to account for minute-by-minute sensitivity changes.

We consider a fixed model for the merger rate evolution that is uniform in luminosity volume. This produces a distribution whose merger rate increases with redshift and is consistent with past estimations~\citep{Nitz:2021zwj,KAGRA:2021duu}, and is motivated by the expected $\rho^{-3}$ falloff for the cumulative distribution of a uniformly distributed source population, as seen in Fig.~\ref{fig:snrfit}. Although we do not consider it here, exploring and constraining alternate redshift evolution models could provide further insight into the origins of the observed black hole population. 

%Fig.~\ref{fig:vt} shows the surveyed volume-time as a function of source-frame chirp mass for O1-O3, O4, and the combined dataset. Although slight, the increase in sensitivity of O4 relative to past observing runs shifts the peak in the sensitive volume-time distribution towards lower masses.

Following a similar approach to~\cite{Nitz:2021zwj}, we assume a mass distribution that is a sum of the estimates from each observation. The merger rate is then obtained by assuming that the number of observations follows a Poisson distribution with a Jeffreys prior on the rate parameter. We marginalize over the statistical uncertainties in the mass distribution caused by the uncertainties in individual event mass estimates. The resulting merger rate estimates (at z=0) for the overall population and individual sub-selections are shown in Table~\ref{tab:merger_rates}.

\begin{table}[ht]
\centering
\caption{The observed merger rate for different chirp mass ranges at $90\%$ confidence.}
\label{tab:merger_rates}
\begin{tabular}{lcc}
\hline
\textbf{Chirp Mass ($M_{\odot}$)} & \textbf{\#} & \textbf{Rate (Gpc$^{-3}$ yr$^{-1}$)} \\
\hline
$[1, 1.5]$ & 2 & $56^{+99}_{-40}$ \\
$[1.5, 3.5]$ & 5 & $36^{+32}_{-20}$ \\
$[3.5, 100]$ & 206 & $19.3^{+3.8}_{-2.1}$ \\
$[3.5, 10]$ & 47 & $15.3^{+5.6}_{-2.9}$ \\
$[10, 100]$ & 159 & $3.8^{+0.5}_{-0.9}$ \\
$[10, 17]$ & 20 & $1.8^{+0.7}_{-0.7}$ \\
$[17, 50]$ & 132 & $1.8^{+0.5}_{-0.3}$ \\
$[50, 100]$ & 7 & $0.015^{+0.014}_{-0.009}$ \\
\hline
\end{tabular}
\end{table}

\section{Conclusions}

In this work, we have presented a preliminary catalog and population analysis of compact binary merger candidates from the ongoing fourth observing run of the LIGO-Virgo-KAGRA network. By analyzing public alerts, we have provided timely estimates of the source-frame chirp mass for over 200 new binary black hole candidates, substantially increasing the size of the known gravitational-wave catalog. Our analysis confirms that the O4 population is broadly consistent with that observed in previous runs, while providing a much sharper view of the features in the black hole mass spectrum. We find continued evidence for a complex distribution with dominant peaks near $8\,\m$ and $30\,\m$, along with continued support for an intermediate feature, lending credence to models that invoke multiple formation channels, including hierarchical mergers in dense stellar environments.

The O4 data has already yielded remarkable individual events, such as the exceptionally loud candidate GW250114, and has added new candidates to the upper mass gap, further challenging our understanding of stellar evolution. While this analysis is preliminary and subject to the systematics discussed, it demonstrates the significant scientific value of low-latency data products. The full catalog of O4 candidates and their estimated properties presented in this work are available in the corresponding data release~\citep{o4popdatarelease}. This release also includes the analysis scripts used to produce the key results and figures in this paper.

\section*{Acknowledgments}
AHN, KK, and KS acknowledge support from the NSF grant PHY-2309240. CDC acknowledges support from the NSF grant PHY-2309356. We acknowledge the support of Syracuse University for providing the computational resources through the OrangeGrid High Throughput Computing (HTC) cluster supported by the NSF award ACI-1341006.

This research has made use of data or software obtained from the Gravitational Wave Open Science Center (gwosc.org), a service of the LIGO Scientific Collaboration, the Virgo Collaboration, and KAGRA. This material is based upon work supported by NSF's LIGO Laboratory which is a major facility fully funded by the National Science Foundation, as well as the Science and Technology Facilities Council (STFC) of the United Kingdom, the Max-Planck-Society (MPS), and the State of Niedersachsen/Germany for support of the construction of Advanced LIGO and construction and operation of the GEO600 detector. Additional support for Advanced LIGO was provided by the Australian Research Council. Virgo is funded, through the European Gravitational Observatory (EGO), by the French Centre National de Recherche Scientifique (CNRS), the Italian Istituto Nazionale di Fisica Nucleare (INFN) and the Dutch Nikhef, with contributions by institutions from Belgium, Germany, Greece, Hungary, Ireland, Japan, Monaco, Poland, Portugal, Spain. KAGRA is supported by Ministry of Education, Culture, Sports, Science and Technology (MEXT), Japan Society for the Promotion of Science (JSPS) in Japan; National Research Foundation (NRF) and Ministry of Science and ICT (MSIT) in Korea; Academia Sinica (AS) and National Science and Technology Council (NSTC) in Taiwan.

\bibliography{sample631}{}
\bibliographystyle{aasjournal}

\end{document}